# Measuring the number of hadronic jets

F.V. Tkachov*[a]

*Physics Department, Penn State University, State College, Pennsylvania 16802*

A quantitative description of the qualitative feature of multihadron final states known as the "number of jets" is given by a sequence of infrared finite shape observables (jet discriminators) that: take continuous values between 0 and 1; are stable—unlike clustering algorithms—against small variations of the input (data errors, Sudakov effects etc.); have a form of multiparticle correlators that is natural in the context of quantum field theory and hence are better suited for a systematic study of theoretical uncertainties (logarithmic and power corrections).

1. The jet paradigm is the foundation of the high-energy collider physics[1]. It is based on the experimental evidence for hadronic jets[2] and the Quantum Chromodynamics-based picture of hadronic energy flow inheriting the shape of partonic energy flow in the underlying hard process[3]. However, the problem of adequate numerical description of multijet structure of multihadron events proved theoretically subtle, its apparent simplicity turned out deceptive, while its satisfactory solution, elusive. The fundamental role of calorimetric measurements in high-energy experiments warrants a scrutiny of the logical principles of such measurements.

2. It makes sense to divide the problems where jets are studied into two classes. The *descriptive theory of hadronic jets* studies the dynamics of jets as such[4]; one is mostly interested in qualitative effects that occur in the leading logarithmic order; a systematic improvement of theoretical predictions is, typically, hardly possible.[5]

The second class (*precision measurements*) comprises quantitative studies of the Standard Model (determination of $\overline{\alpha}_S(Q^2) \to \infty$ etc.[1]) where one aims at a highest reliability for both data and theoretical predictions:

Reliability of data means that the problem ought to be regarded as the one of measurement rather than the one of modeling dynamics. One has to ensure that measurements be stable with respect to errors in data from calorimeter cells, their position and geometry, etc. (otherwise physical information may be distorted by artefacts of measurement procedures), and that the data experimentalists produce be not biased by the imperfect knowledge of details of dynamics.

Reliability of theoretical predictions means that it ought to be possible to systematically include logarithmic and power corrections. The observables one uses ought to conform to the general structure of the underlying formalism (perturbative Quantum Field Theory) to ensure a better control over theoretical uncertainties due to a considerable sophistication of the modern analytical methods of the theory of Feynman diagrams[6].

3. Jet counting is an attempt to use jets of hadrons to tag events. Its great usefulness[7] is due to the fact that the very presence of jets and their number is the most direct and clear manifestation of the dynamics of QCD.

The conventional jet counting determines an integer number of jets for each event using algorithms[8] which attempt to reconstruct the underlying partons' momenta by, in effect, inverting the hadronization. They were invented[9] in the context of descriptive theory of jets, involve many ambiguities[8], and their use in measurement-type problems may not be accepted uncritically:

On the theory side, the definition of jets in such algorithms uses phase space cutoffs to take into account cancellations of IR singularities. This is rather unnatural within the formalism of QFT: one has to recur to numerics even in simpler cases[10] whereas a study of power corrections remains practically impossible.

On the measurement side, *any* algorithm that produces an *integer* number of jets cannot be fully satisfactory— even before any dynamics gets involved. Indeed, such an algorithm rips the continuum of multiparticle states by mapping it to the discrete set of natural numbers. A discontinuous mapping is unstable with respect to small variations (measurement errors or unknown higher order corrections) for some values of input data (cf. Fig. 1).[11] As a result, the inversion of hadronization is a mathematically ill-posed problem, whence the problem of spurious jets, and the sensitivity to Sudakov effects and to irrelevant details of recombination procedures.[8] This pathology is somewhat masked off by averaging over many events. But a deterministic recombination algorithm is applied separately to each stochastically generated event. So, instead of a statistical compensation of errors, there occurs a smearing between cross sections with adjacent numbers of jets.[12] It can be eliminated neither by increasing statistics, nor by varying the jet resolution $y_{cut}$, and it is more important for smaller $y_{cut}$, lower energies and larger numbers of jets.[13]

What, then, could be a quantitative measure for the qualitative feature of multiparticle final states known as the "number of jets", a measure that allows a correct handling of data errors and a systematic study of theoretical uncertainties, and that is unbiased by the imperfect knowledge of jet dynamics?

4. MATHEMATICAL NATURE OF "ENERGY FLOW". If $\omega$ is a calorimeter cell, then the energy deposited in it by particles that hit the cell is $E(\omega) \geq 0$. Energy conservation implies that if one takes two non-overlapping cells $\omega$ and $\omega'$ and combines them into one, then the energy deposited in it is the sum of energies deposited in $\omega$ and $\omega'$ separately: $E(\omega \cup \omega') = E(\omega) + E(\omega')$. One can consider cells $\omega$ simply as parts (subsets) of the unit sphere around the collision point. Then the energy flow (EF for short) is a non-negative additive function on the subsets $\omega$. Such functions are known as *abstract measures*.[14, 15]

Let $P$ be a multiparticle state, $P = \{E_i, \hat{p}_i\}_i$, where $E_i$ and $\hat{p}_i$ are the energy and direction (a unit 3-vector) of the $i$-th particle. All information about $P$ obtainable using calorimeters is its EF represented as a linear combination of $\delta$-functions localized at $\hat{p}_i$,

$$E_P(\hat{p}) = \sum_i E_i \delta_{\hat{p}_i}(\hat{p}) \ , \qquad (1)$$

where $\hat{p}$ is a variable unit 3-vector running over the sphere. The energy measured by a cell $\omega$ is

$$E_P(\omega) = \int_\omega d\hat{p} \ E_P(\hat{p}) = \sum_{\hat{p}_i \in \omega} E_i \ .$$

The observables we deal with in calorimetric measurements are functions of EFs $E$. Let $f(E)$ be such a function. Its stability with respect to data errors translates into a concrete kind of continuity. Let $E_n$ be a sequence of EFs such that, however small the energy resolution and geometry of calorimeters, $E_n$ become indistinguishable within data errors for all $n$ large enough. This *calorimetric* or *C-convergence* of EFs is formalized as follows. Let $0 \le \varphi(\hat{p}) \le 1$ be a continuous function on the unit sphere. It can be thought of as describing local efficiency of a calorimeter cell: for a given EF $\boldsymbol{E}(\hat{p})$ the expression $\int d\hat{p} \, \boldsymbol{E}(\hat{p}) \varphi(\hat{p}) \equiv \langle \boldsymbol{E}\varphi \rangle$ is the energy measured by this cell. Then *C-convergence* of $\boldsymbol{E}_n$ is equivalent to numerical convergence of $\langle \boldsymbol{E}_n \varphi \rangle$ for any "detector" $\varphi$.[16] For a correctly defined observable $f$, $f(\boldsymbol{E}_n)$ should converge in numerical sense for any such sequence $\boldsymbol{E}_n$. Such functions $f$ (*calorimetric*, or *C-observables*) are exactly the ones that are stable with respect to measurement errors of calorimetric detectors.

5. The role of *C-continuity* is best understood with the help of analogy with the familiar length measurements. Length is habitually represented as a real number—an idealization that, one tends to forget, is highly non-trivial from a historical perspective. In particular, the familiar continuity of real numbers is useful *only* inasmuch as it corresponds to the stability of, say, volume computations with respect to data errors of the length measurements involved. The elusive reality of calorimetric measurements is that whereas rulers measure length as a real number, calorimetric detectors measure energy flow as an additive function on subsets, and the *C-continuity* plays exactly the same role for data errors of calorimeters as the usual continuity of real numbers does for rulers.

6. A large class of *C-observables* is immediately found as follows.[17] Consider the direct product of $m$ identical EFs $\boldsymbol{E}(\hat{p})$. Then the standard theorems[14] imply *C-continuity* of the observables of the following form:

$$F_m(\boldsymbol{E}) = \int d\hat{p}_1 \ldots \int d\hat{p}_m \, \boldsymbol{E}(\hat{p}_1) \ldots \boldsymbol{E}(\hat{p}_m) \, f_m(\hat{p}_1, \ldots, \hat{p}_m) \; , \quad (2)$$

where $f_m$ is any continuous symmetric function. A function on EFs induces a function on multiparticle states: using (1) and (2) one obtains:

$$F_m(\{E_i, \hat{p}_i\}) = \sum_{i_1 \ldots i_m} E_{i_1} \ldots E_{i_m} \, f_m(\hat{p}_{i_1}, \ldots, \hat{p}_{i_m}) \; . \quad (3)$$

This is automatically fragmentation invariant. If $f_m$ satisfy minimal requirements of regularity (e.g. existence of first derivatives), then $F_m$ are IR finite[18]. Such *C-observables* are interpreted as average values of operators that are $m$-local in momentum space, which offers a possibility of their systematic theoretical study.[19] Examples of *C-observables* are the well-known thrust, sphericity, etc. (see Ref. 4 for a complete list).[20]

Algebraic combinations of *C-observables* are again *C-observables*. But taking e.g. infinite sums of such functions ($m \to +\infty$) requires care: one can arrive at observables that are IR safe in each order of perturbation theory, continuous in the ordinary sense as functions of particles' energies and momenta for any fixed number of particles, but not *C-continuous*.[21] A complete understanding of this subtlety in a general QFT context is lacking. Anyhow, *C-continuity* limits available options, and if one wishes to deal with correlator-type observables then the above $F_m$ remain the only choice.

7. Measuring the "number of jets". Imagine a step function equal to 0 on states with less than $m$ jets, and to 1 elsewhere. A sequence of such functions ($m = 1, 2, \ldots$) would do the job of jet counting just fine. But we wish to deal with *C-observables*. So, consider a *C-observable* (3) that is exactly 0 on any state with less than $m$ particles. Then $f_m(\hat{p}, \hat{p}, \ldots) = 0$, so that $f_m(\hat{p}_1, \hat{p}_2, \ldots)$ should contain a nullifying factor $\Delta_{12}$, e.g.:

$$\Delta_{12} = 1 - \cos\theta_{12} = (p_1 p_2)(p_1 P)^{-1}(p_2 P)^{-1} \; , \quad (4)$$

where $p_i$ are light-like 4-momenta ($p_i^0 = E_i$, $\vec{p}_i = E_i \hat{p}_i$) and the 4-vector $P^2 = 1$ describes the reference frame[22]. Symmetry yields a similar factor for each pair of arguments. One gets the sequence of *jet discriminators*:

$$J_m(E_1, \hat{p}_1, \ldots, E_N, \hat{p}_N) = \sum_{1 \le i_1 < \ldots < i_m \le N} E_{i_1} \ldots E_{i_m} \, j_m(\hat{p}_{i_1}, \ldots, \hat{p}_{i_m}) \; ,$$

$$j_m(\hat{p}_1, \ldots, \hat{p}_m) = N_m \prod_{1 \le i < j \le m} \Delta_{ij} \; . \quad (5)$$

It turns out that $0 \le J_m \le 1$ on any state if $N_m$ is defined from the condition $J_m(\boldsymbol{P}_\infty^{\text{sym}}) = 1$ where $\boldsymbol{P}_\infty^{\text{sym}}$ is a limit of uniformly distributed states $\boldsymbol{P}_N^{\text{sym}}$ with $N \to \infty$ so that $E_i = N^{-1}$ and $\sum_i \to \frac{N}{4\pi} \int d\hat{p}_i$. Then $N_2 = 2$, $N_3 = 27/4$, $N_4 = 36$, $N_5 = 9375/32$, $N_6 = 455625/128$.[23]

Some special values of $J_m(\boldsymbol{P})$ are as follows. For the state $\boldsymbol{P}_3^{\text{sym}}$ consisting of 3 symmetrically arranged particles, $J_3(\boldsymbol{P}_3^{\text{sym}}) = 27/32 \approx 0.84$. For a symmetric state of 4 particles (tetrahedron) $J_3(\boldsymbol{P}_4^{\text{sym}}) = 1$, $J_4(\boldsymbol{P}_4^{\text{sym}}) = 64/81$. For a symmetric state of six particles (octahedron), $J_3(\boldsymbol{P}_6^{\text{sym}}) = J_4(\boldsymbol{P}_6^{\text{sym}}) = 1$.

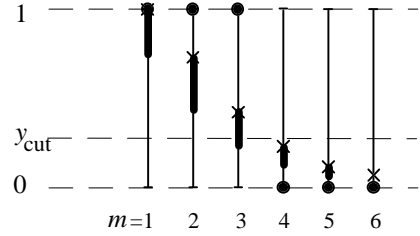

Fig. 1. The crosses are the values of the jet discriminators $J_m$ for a typical final state. When looked at sideways, the fat lines represent the "number of jets" as a function of $y_{\text{cut}}$.

Fig. 1 shows a typical picture of values of jet discriminators. The usual jet counting amounts to replacing the continuous $J_m$ with 0 or 1 (the blobs). This can be achieved e.g. by introducing a cutoff $y_{\text{cut}}$ as shown.[24] The non-zero tail at large $m$ is due to hadronization (including Sudakov effects). The instability with respect to such effects as well as to data errors (shifts of the crosses) is clearly seen.[25]

Note that $J_m \equiv 0$ for $m$ larger than the number of particles (or detector cells). For decreasing width $\Theta$ of jets and for $m > M$ (a typical number of jets in the event), $J_m$ are increasingly suppressed by powers of $\Theta^2$ and of the energy fractions of soft particles. This ensures a monotonic decrease of the values of $J_m$ for $m > M$ for typical events. Numerical experiments show that the decrease of $J_m$ is a universal feature even for $m \lesssim M$.[23,26]

Fragmentation causes the values of $J_m$ to increase as compared with the parton state. However, the *C-continuity* of $J_m$ ensures that the closer (in the calorimetric sense) the final hadron state is to the parton state, the less the difference in values of *C-observables*, and the less the upward shift of $J_m$.

8. For the case of hadrons in the initial state one should modify $J_m$ to suppress contributions from the hadron beams. For $pp$, say, it is sufficient to introduce into $j_m$ the factor $1 - \cos^2\theta_i$ per each particle where $\theta_i$ is the angle between the particle's direction and the beam axis.

9. Now, fix a multiparticle state $P$ and consider any jet counting algorithm $A$ that produces an integer number of jets $N_A(y_{cut}; P)$ for each $y_{cut}$, which is non-decreasing as $y_{cut} \to 0$. Then from Fig. 1 one sees that one could, in theory, restore a sequence of jet discriminators $J_m^A(P)$ similar to $J_m(P)$. Thus, the information content of $J_m^A(P)$ and $N_A(y_{cut}; P)$ is essentially equivalent. But it is hardly possible to find meaningful expressions for $J_m^A(P)$ for the popular algorithms. Our $J_m(P)$ are singled out by the transparency of analytical structure.

10. So, studying the average values of jet discriminators $\langle J_m \rangle$ (qualitatively interpreted as fractions of events with no less than $m$ jets) instead of the usual $n$-jet fractions may have an advantage of reducing, in perspective, both theoretical and experimental uncertainties.

To compute $\langle J_m \rangle$ from data, one would treat each calorimeter cell as a particle (the correctness of this is ensured by $C$-continuity). Computations can be optimized due to the regular structure of $J_m$ in several ways: (i) One can do the summations à la Monte-Carlo with probabilities equal to energy fractions. (ii) A preclustering can be used due to $C$-continuity to reduce the number of particles to, say, $\leq 30$ when computations are easily manageable; since the exact expression is known, the approximation errors are fully under control here. (iii) The computation of (5) can be parallelized.

On the theoretical side, studying effects of hadronization would reduce to studying logarithmic and power corrections to $\langle J_m \rangle$. Resummation of logarithms is done via the standard renormalization group. The analytical calculations of the corresponding diagrams are easier due to the simple analytical form of the weights in the phase space integrals (cf. (4)). Also, a prospect opens for a study of power corrections[27]. Recall that the power corrections for $\sigma_{tot}(e^+e^- \to hadrons) \propto J_2$ are given by expressions involving vacuum condensates[28] that are directly related to soft singularities; the structure of power corrections can be obtained within perturbation theory[29] while the values of condensates are estimated via the lattice QCD. A similar approach should be feasible for the jet discriminators.[30]

11. The importance of the problem of jet definition was impressed upon me by S. D. Ellis. A crucial encouragement came from A. V. Radyushkin. I thank Z. Kunszt and ETH for the hospitality during the Workshop on New Techniques for Calculating Higher Order QCD Corrections (ETH, Zürich, December 1992)—its atmosphere catalyzed the present work. Yu. Bashmakov, S. Catani, R. K. Ellis, I. F. Ginzburg, B. L. Ioffe, A. Klatchko, Z. Kunszt, D. V. Shirkov, T. Sjöstrand, I. K. Sobolev, B. Straub, S. Youssef and the three referees supplied bibliography and/or necessary criticisms. I thank the participants of several workshops and seminars for lively discussions, and J. C. Collins and H. Grotch for the hospitality at the Penn State University where this work was completed. It was supported in parts by the U.S. Department of Energy (grant DE-FG02-90ER-40577) and by the International Science Foundation (grant MP9000).

*     On leave of absence from the Institute for Nuclear Research of Russian Academy of Sciences, Moscow 117312, Russia.

[1] For a review see S. D. Ellis, "Lectures on perturbative QCD, jets and the Standard Model: collider phenomenology" at the 1987 Theoretical Advanced Study Institute (1987, July, St. John's College, Santa Fe, NM), available as technical report NSF-ITP-88-55.

[2] G. Hanson et al., Phys. Rev. Lett. **35**, 1609 (1975).

[3] G. Sterman and S. Weinberg, Phys. Rev. Lett. **39**, 1436 (1977).

[4] R. Barlow, Rep. Prog. Phys. **36**, 1067 (1993).

[5] For instance, one needs to identify jet axes to study the QCD coherence (Yu. L. Dokshitzer et al., Rev. Mod. Phys. **60**, 373 (1988)), but the notion of jet's axis is ambiguous beyond the leading order of perturbation theory.

[6] Cf. F. V. Tkachov, in Ref. 7.

[7] Cf. Proc. New Techniques for Calculating Higher Order QCD Corrections (ETH, Zürich, 16-18 December 1992), ed. Z. Kunszt (ETH, Zürich, 1992).

[8] For a review see S. Bethke et al., Nucl. Phys. **B370**, 310 (1992) and S. Catani, in: Proc. 17th INFN Eloisatron Project Workshop, Erice, 1991, ed. L. Cifarelli and Yu. L. Dokshitzer (Plenum Press, New York).

[9] T. Sjöstrand, Comp. Phys. Commun. **28**, 229 (1983); see also Ref. 8.

[10] Cf. W. T. Giele and E. W. N. Glover, Phys. Rev. **D46**, 198 (1991).

[11] A more stable variant of recombination was described in S. Youssef, Comp. Phys. Commun. **45**, 423 (1987).

[12] Cross sections for adjacent numbers of jets differ by $O(\alpha_s)$, so 1% of 3-jet events identified as having 4 jets (due to data errors or incomplete knowledge of hadronization) means an $O(10\%)$ error for the 4-jet cross section.

[13] Multijet channels can be used e.g. for top search (F. A. Berends et al., Nucl. Phys. **B357**, 32 (1991)). On the other hand, the study of multijet cross sections by UA2 (K. Jacobs, in "Joint Int. Lepton-Photon Symp. on HEP" (1991, Geneva) World Scientific: Singapore, 1992) concluded that the agreement of the data with theory they found for 4–6 jets "can be considered as largely accidental".

[14] For a thorough treatment see L. Schwartz, Analyse Mathématique, vol. 1 (Hermann, Paris, 1967).

[15] The word "measure" is used here in two different meanings, physical and mathematical, not to be confused.

[16] Note that all possible EFs form an infinitely-dimensional space, and there are many radically non-equivalent ways to define convergence (i.e. topology) in such spaces (see any textbook on functional analysis). To appreciate the subtlety of the problem recall the large-scale study of G. C. Fox and S. Wolfram, Nucl. Phys. **B149**, 413 (1979), who adopted an incorrect idealization of EFs as functions on the unit sphere with $L_2$ topology (familiar from quantum mechanics) and the research was lead astray towards studying spherical harmonics etc. Our $C$-convergence is the well-known weak convergence of linear functionals. It cannot be described in terms of a single-valued distance function or norm (cf. Ref. 14). This may be psychologically uncomfortable but such is the nature of calorimetric measurements.

[17] Cf. the cumbersome constructions of Fox and Wolfram, in Ref. 16.

[18] G. Sterman, Phys. Rev. **D19**, 3135 (1979).

[19] Cf. F. R. Ore and G. Sterman, Nucl. Phys. **B165**, 93 (1980).

[20] Note that the energy correlations of C. C. Basham et al., Phys. Rev. Lett. **41**, 1585 (1978) are a special case of $F_m$—but with a discontinuous $f_m$. Since hadronization is not inverted here, the lack of $C$-continuity is numerically less important than with recombination algorithms.

[21] F. V. Tkachov, in "Joint International Workshop on High Energy Physics" (September 1993, Zvenigorod, Russia), ed. B. B. Levtchenko (INP MSU: Moscow, 1994), p. 80.

[22] For $e^+e^-$ this is the total 4-momentum. For $ep$ one may wish to choose a different frame—as described e.g. by B. Webber, J. Phys. **G19**, 1567 (1993).

[23] I thank B. B. Levtchenko for numerical checks of this.

[24] Such a jet counting procedure was called Moscow sieve in F.V. Tkachov, technical report, INR-F5T/93-01 (unpublished).

[25] The figure suggests that the importance of Sudakov effects in the conventional recombination algorithms is an artefact due to the instability of the latter.

[26] S. Catani et al., Nucl. Phys. B406, 187 (1993) discuss relationship between jet clustering algorithms and event shape measures, including the monotonicity.

[27] whose importance is discussed by B. W. Webber, in Ref. 7.

[28] M. A. Shifman et al, Nucl. Phys. **B147**, 448 (1979).

[29] F. V. Tkachov, Phys. Lett. **B125**, 85 (1983).

[30] The required mathematics is reviewed in Ref. 6.